\documentclass[conference, 10pt]{IEEEtran}

\usepackage{amssymb} 

\usepackage[T1]{fontenc}
\usepackage[english]{babel}
\usepackage[normalem]{ulem}
\usepackage{algorithm2e} 
\usepackage{mathtools} 
\usepackage{siunitx}
\usepackage{tikz}

\usepackage{fancyhdr} 
\usepackage{setspace}
\usepackage{ifthen} 
\usepackage[utf8]{inputenc} 

\definecolor{Yellow}{rgb}{1,0.9,0.7}
\definecolor{Pink}{rgb}{1,0.85,0.85}
\definecolor{AntiqueWhite}{rgb}{0.9,0.9,0.9}

\newcommand{\NOTE}[1]%
{
\noindent
\fcolorbox{black}{AntiqueWhite}{\parbox{0.95\columnwidth}}
{\textbf{NOTE: } #1}
}

\begin{document}
\title{Characterizing Accuracy Trade-offs of EEG Applications on Embedded HMPs}


\author{\IEEEauthorblockN{Zain Taufique}
\IEEEauthorblockA{\textit{University of Turku}\\
zatauf@utu.fi}
\and
\IEEEauthorblockN{Muhammad Awais Bin Altaf}
\IEEEauthorblockA{\textit{Lahore University of Management and Technology}\\
awaisbinaltaf@gmail.com}
\and
\IEEEauthorblockN{Antonio Miele}
\IEEEauthorblockA{\textit{Politecnico di Milano}\\
antonio.miele@polimi.it}
\and
\IEEEauthorblockN{Pasi Liljeberg}
\IEEEauthorblockA{\textit{University of Turku}\\
pasi.liljeberg@utu.fi}
\and
\IEEEauthorblockN{Anil Kanduri}
\IEEEauthorblockA{\textit{University of Turku}\\
spakan@utu.fi}
}

\date{}
\maketitle

\thispagestyle{empty}
\begin{abstract}
Electroencephalography (EEG) recordings are analyzed using battery-powered wearable devices to monitor brain activities and neurological disorders. These applications require long and continuous processing to generate feasible results. However, wearable devices are constrained with limited energy and computation resources, owing to their small sizes for practical use cases. Embedded heterogeneous multi-core platforms (HMPs) can provide better performance within limited energy budgets for EEG applications. Error resilience of the EEG application pipeline can be exploited further to maximize the performance and energy gains with HMPs. However, disciplined tuning of approximation on embedded HMPs requires a thorough exploration of the accuracy-performance-power trade-off space. In this work, we characterize the error resilience of three EEG applications, including Epileptic Seizure Detection, Sleep Stage Classification, and Stress Detection on the real-world embedded HMP test-bed of the Odroid XU3 platform. We present a combinatorial evaluation of power-performance-accuracy trade-offs of EEG applications at different approximation, power, and performance levels to provide insights into the disciplined tuning of approximation in EEG applications on embedded platforms. 
\end{abstract}
\section{Introduction}
Electroencephalography (EEG) recordings are commonly used for diverse applications including emotion detection \cite{emotion2}, sleep stage classification \cite{sleep1}, \textit{epileptic seizure prediction}\cite{zain-epi}, \textit{migraine detection} \cite{zain-mig} etc. EEG electrodes are placed on the scalp targeting specific areas of the brain, using the 10-20 electrodes system \cite{EEG-10-20}. Wearable devices are used to record analog EEG signals from different channels and pre-process raw inputs -- to remove noise, convert the analog signals to digital, and extract application-specific features from the digital signals. EEG applications are time-critical and require continuous input data processing to produce feasible results~\cite{Ear-EEG}. 
On the other hand, wearable devices can offer only minimal performance, within the limited compute capacity, and energy budgets \cite{zain-epi}. Given these resource constraints, wearable devices offload compute-intensive tasks onto nearby edge devices, to meet application-level performance requirements ~\cite{Task-Offload, Ometov-Offload}. The approximation is another alternative strategy to meet performance requirements within energy constraints, exploiting the accuracy-performance trade-offs in applications with inherent error resilience~\cite{Anil-DAC}. Combining computational offloading with approximation significantly improves the efficiency of edge devices, while catering to the performance requirements of time-critical EEG applications, within the energy bounds of wearable devices \cite{shahhosseini2022online}.  
Embedded mobile platforms featuring heterogeneous multi-core processors (HMPs) are widely used as edge devices, specifically in the context of EEG applications \cite{EEG-Multi}. HMP-based edge devices have a wider range of power-performance configuration space, considering tunable knobs such as core-level heterogeneity, application-level parallelism, dynamic voltage, and frequency scaling, and CPU quota allocation \cite{shamsa2021concurrent}. Analogously, application-level approximation exhibits a wider configuration space of accuracy-performance trade-offs, with different levels, and types of approximations \cite{Anil-DAC}. Combining the system-level power-performance trade-offs, and application-level accuracy-performance trade-offs exposes a significantly wider Pareto-optimal space of different power-performance-accuracy levels. The choice of an optimal execution configuration tuple consisting of power-performance-accuracy level operating points requires an exhaustive exploration of the Pareto-optimal space. Further, factors such as core-level heterogeneity, application-specific sensitivity to error, and choice of a dynamic combination of different power knobs along with approximation create a trade-off space with non-linear gains and non-intuitive caveats. This exacerbates the challenge of selecting a tuple of optimal operating points of power-performance-accuracy levels. In this work, we address this challenge through design space exploration to alleviate navigating through the power-performance-accuracy Pareto-optimal space in EEG applications. 
We characterize the performance, and power gains of EEG applications with approximation, executed on embedded HMP-based edge devices. We analyze the 3-d Pareto-optimal space of power-performance-accuracy trade-offs in EEG applications at multiple approximation levels, under different power, and performance knob settings. We present an understanding of error resilience characteristics of EEG applications on embedded HMPs and discuss caveats of exploring accuracy trade-offs for performance, and power gains. In Section \ref{sec.Appx}, we discuss the different types of approximations in EEG applications that can ensure performance gains against accuracy trade-offs. In Section \ref{sec.res}, we discuss our benchmark EEG applications, hardware setup, and characterization results.

\section{Error resilience in EEG Applications} \label{sec.Appx}
Fast Fourier Transform (FFT) is widely used in feature extraction for EEG applications \cite{zain-mig, DOA, zain-mig2}. However, FFT cannot be directly used in this process because FFT is suitable for stationery signals, while EEG signals are non-stationary ~\cite{Nonstationary-EEG}. The \textit{Welch} method was proposed to resolve this problem, where fixed-sized windows of the input signal are processed through the FFT block for feature extraction ~\cite{Welch}.
\subsection{Welch Method}
In this method, we calculate the \textit{PSD} at different frequency bands using the Weighted Overlapped Segment Averaging (WOSA) method. The range of these frequency bands is application-specific. The sliding window mechanism covers all the input samples, where each subsequent window overlaps with the previous window. The PSD is calculated using the formula shown in equation \ref{eqs.PSD}.
\vspace{-3pt}
\begin{equation}
P[n] = \frac{1}{L}\sum_{n=1}^{L} [X_{bp}[n]]^2
\label{eqs.PSD}
\vspace{-3pt}
\end{equation}

\subsection{Approximation Methods} \label{sec.Apx}
The fine-grained application level approximation in the Welch method can be achieved by i) loop perforation, ii) changing the FFT length, and iii) changing the window overlapping. 
In loop perforation, we skip loop iterations to trade-off performance vs. accuracy loss ~\cite{loop-perf}. The behavior of EEG signals can vary for subjects with different age groups, genders, and biological backgrounds. Hence, approximation using loop perforation can only be useful for a bounded data set. Otherwise, randomly skipping loops can lead to the risk of losing critical data, causing uncertainty in the final results. 
Secondly, the FFT used in the Welch method has a defined input length following the Nyquist phenomenon. Increasing the FFT length increases computation load, which impacts the overall power requirements for a resource constraint device. In contrast, decreasing the FFT length reduces the output granularity, impacting output accuracy.
\begin{figure}[t] \vspace{-5pt}
\centering
\includegraphics[width=0.4\textwidth]{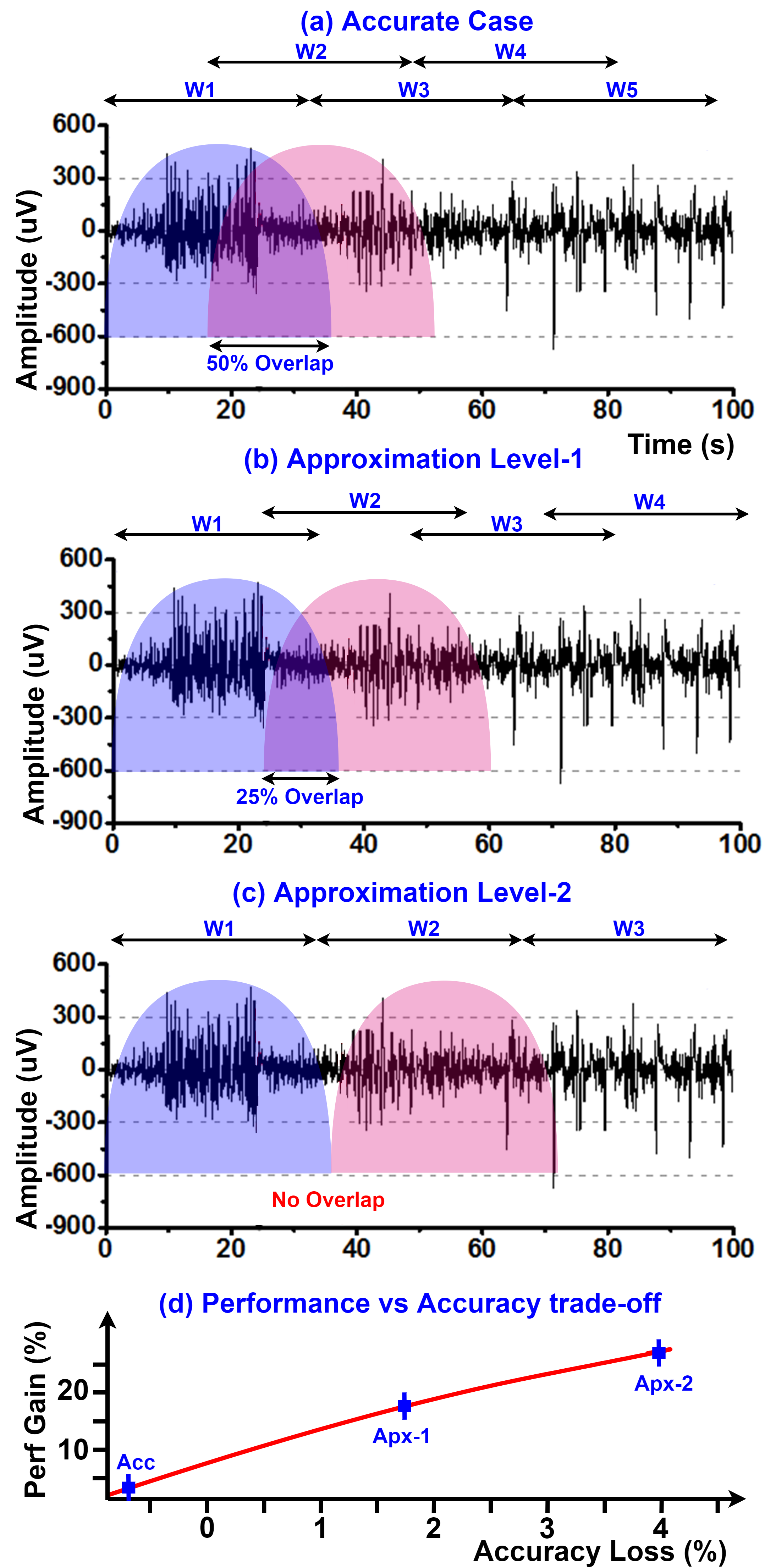}
\caption{The Performance gain vs accuracy loss in Welch using window overlap method (a) Accurate, 50\% overlap, (b) Level-1 approximation, 25\% overlap, (c) Level-2 approximation, no overlap, (d) Performance gain vs Approximation}
\vspace{-12pt}
\label{fig.Apx}
\end{figure}
Finally, the Welch method can also be approximated by changing the overlap percentage of subsequent windows. A typical window of samples is designed to have smooth edges. Each subsequent window overlaps with the previous window at a certain percentage to avoid data loss at the edges. Increasing the percentage overlap increases the number of windows covering the maximum input data samples, decreasing the output latency. The application performance can be enhanced by decreasing the percentage overlap while trading off bounded accuracy loss.
Figure \ref{fig.Apx} shows an example of approximation using window overlapping in the Welch method for Epileptic seizure detection. The EEG signals are taken from the F7-T7 channel recording of an epileptic patient available in the open source \textit{Physionet} CHB-MIT database ~\cite{CHB_MIT}. In Figure \ref{fig.Apx} (a) the percentage overlap between two consecutive windows is 50\%, presenting the accurate case using 5 windows. Figure \ref{fig.Apx} (b) presents the first level of approximation where the percentage overlap is reduced to 25\%, and the number of required windows is reduced to 4. Similarly, Figure \ref{fig.Apx} (c) shows the third level of approximation where there is no overlap between subsequent windows and only 3 windows are required to calculate the output. \ref{fig.Apx} (d) provides insights into the percentage performance gain against each level of approximation. The performance gain is significantly higher than the accuracy loss at each level. 

\section{Characterizing Accuracy Trade-offs} \label{sec.res}
To characterize EEG-based applications on embedded hardware, we require i) EEG data sets, ii) a baseline embedded platform, and iii) an implementation strategy to evaluate the results.
\subsection{Datasets of EEG Applications}
This paper considers three different EEG applications: sleep state classification, Stress monitoring, and Epileptic Seizure detection ~\cite{zain-epi}. For sleep-state classification, we use the extended sleep-EDF data-set by \textit{Physio-Net}, sampled at 100Hz, for two channels: Fpz-Cz and Pz-Oz ~\cite{sleep-EDF}.
We used EDMAT data-set ~\cite{EDMAT} by \textit{PhysioNet} for stress monitoring application. The EEG data were recorded on 19 different channels. For epileptic seizure detection, we used \textit{CHB-MIT} data-set, where EEG signals were sampled at 256 Hz for 24 patients ~\cite{CHB_MIT}. For the presented work, we used the recording of 5 EEG channels (F7-T7, F8-T8, T7-FT9, FT10-T8, FP1-F3).

\subsection{Hardware Setup}
We used the Odroid XU3 board featuring Samsung Exynos 5422 chip with Linux Ubuntu 15.04 as the baseline heterogeneous multi-core platform. It is based on ARM big.LITTLE architecture with four Cortex-A7 CPU cores on the \textit{LITTLE} cluster, and four Cortex A15 cores on the \textit{big} cluster. This board supports traditional power knobs of Dynamic Voltage and Frequency Scaling (DVFS) and application-to-core mapping. We load Operating system (OS) drivers into the platform to form an interface between the applications and the hardware. We use the DVFS options to adjust the CPU frequency of each cluster, with ranges 200Hz - 1400Hz for the LITTLE  cluster and 200Hz - 2000Hz for the big cluster. For application-to-core mapping, we use the OS interface to map the applications on the required number of cores of a cluster. We also use the in-built power sensors to monitor the current power of each CPU cluster. We have designed a Run-Time Monitor to store the applications' power, performance, and accuracy logs at the settings of the given knobs.

\subsection{Implementation Strategy}
We implemented each EEG application using the Welch method with Bartlett-Hanning window and 1024 point FFT. We designed these applications using pthreads to support parallel workloads on a multi-core platform. After the FFT calculations, we apply bandpass filters according to the target application. Table \ref{Freq-bands} shows the frequency bands used for three different target applications. The PSD calculations on the given frequency bands formulate the feature vector of each application.

\begin{table}[t]
\vspace{-3pt}
\centering
 \caption{Frequency bands (Hz) for Seizure Detection, Sleep Stage Classification, and Stress Detection}
 \label{Freq-bands}
  \scalebox{0.9}{
 \begin{tabular}{|l|c|c|c|c|c|}
 \hline
 \textbf{Application} & \textbf{Delta} & \textbf{Theta} & \textbf{Alpha} & \textbf{Beta} & \textbf{Gamma}\\
 \hline
    Seizure \cite{zain-epi} & 0.5-2 & 2-6 & 6-8 & 8-30 & >30\\
 \hline
    Sleep ~\cite{sleep-Band} & 0.5-3.5 & 3.5-7.5 & 7.5-12 & 12-30 & >31\\
 \hline
    Stress ~\cite{Stress} & 0-3.9 & 4-7.9 & 8-10 & 14-29.9 & 30-47\\
 \hline
 \end{tabular}
 }
\end{table}

\section{Evaluation and Results}
For evaluation, we use offline ANN models to calculate the accuracy of the results as a verification parameter. The accuracy is formulated as a fraction of True Negatives (TN), True Positives (TP), False Negatives (FN), and False Positives (FP) as shown in equation \ref{eq:mAcc}. Where TP represents a calculated positive detection against an actual positive detection, FP represents a calculated positive detection against an actual negative detection, and vice versa.

\vspace{-6pt}
\begin{equation}
    Accuracy = \frac{(TP+TN)}{(TP+TN+FP+FN)}
    \label{eq:mAcc}
\end{equation}

We use the Heartbeat API ~\cite{hbAPI} to calculate the application level performance (\textit{Hb/s}). We compare the application characteristics on three different DVFS settings of 600 Hz, 1000 Hz, and 1400 Hz for each cluster. At each frequency level, we also map the application at a different number of CPU cores. We characterized the applications on five different approximation levels following the window overlap technique explained in Section \ref{sec.Apx}. 
The accurate configuration has a 50\% overlap between two consecutive windows, while at every increasing approximation level, we decrease the percentage overlap by 10\%. We have divided the characterization of EEG applications in terms of three different CPU frequency levels.


\begin{figure*}[t]
\vspace{-3pt}
\centering
\includegraphics[width=0.99\textwidth]{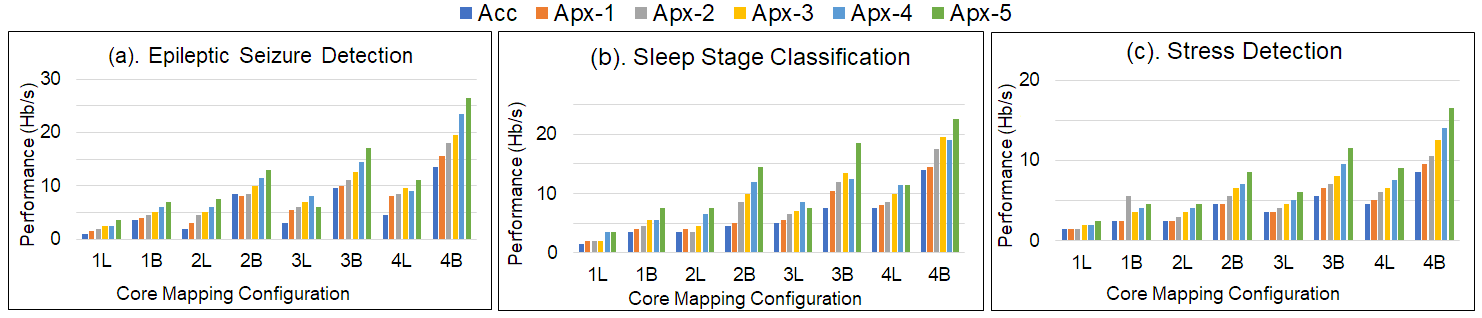}
\caption{Performance characteristics at 600 Hz for (a) Epilepsy, (b) Sleep, (c) Stress.}
\label{fig.Perf_600.Apx}
\vspace{-6pt}
\end{figure*} 

\begin{figure*}[t]
\vspace{-3pt}
\centering
\includegraphics[width=0.99\textwidth]{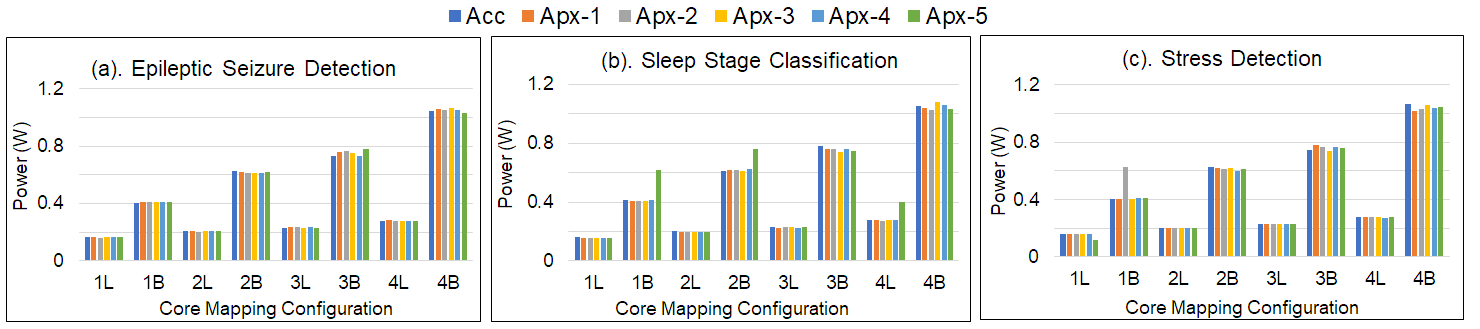}
\caption{Power characteristics at 600 Hz for (a) Epilepsy, (b) Sleep, (c) Stress.}
\label{fig.Power_600.Apx}
\vspace{-6pt}
\end{figure*} 

\begin{figure*}[t]
\vspace{-3pt}
\centering
\includegraphics[width=0.99\textwidth]{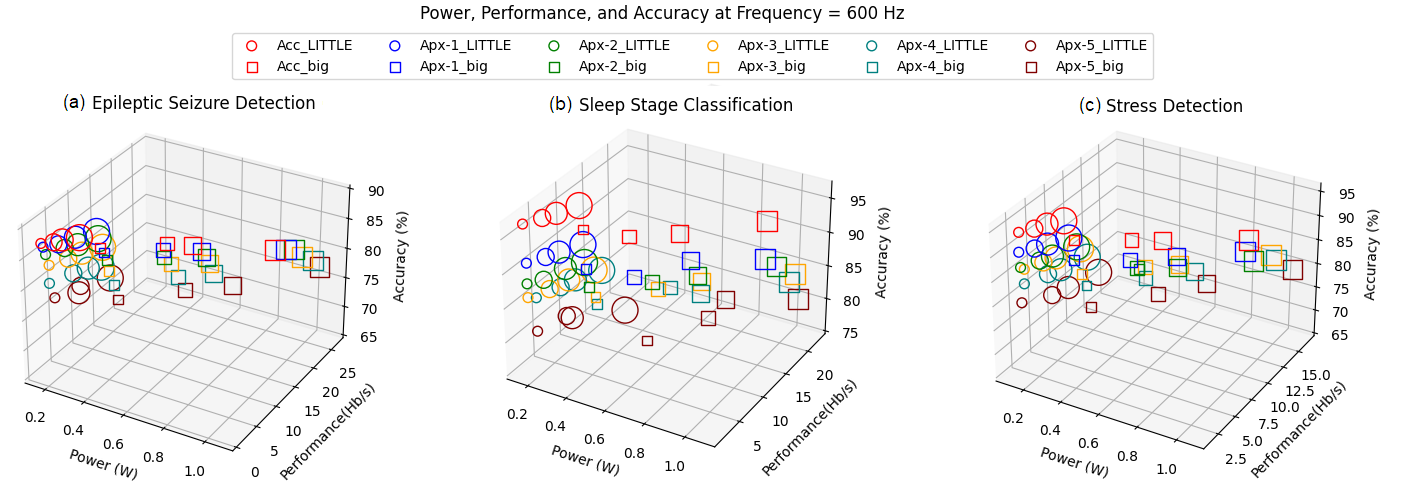}
\caption{Accuracy trade-off vs power-performance characteristics at 600 Hz for (a) Epilepsy, (b) Sleep, (c) Stress.}
\vspace{-15pt}
\label{fig.3d_600.Apx}
\end{figure*}

\textbf{Frequency Level-1.} We ran all three EEG applications on eight different core mapping configurations at 600 Hz CPU frequency levels at both clusters. Figure \ref{fig.Perf_600.Apx} compares the performance of all three applications against accurate and approximate configurations. The general trend shows that increasing the approximation level increases the application performance. Moreover, for the same number of cores on each cluster, we see a better performance at the big luster. Each application performs less when mapped on one LITTLE core and the highest while running on four big cores.
Figure \ref{fig.Power_600.Apx} shows the power trends for all three applications at 600 Hz frequency on both CPU clusters. In general, the applications consume higher power at the big cluster, where the highest running power reaches 1.1 W for almost all three applications. The power consumption trends are somewhat similar between different levels of approximation for each core mapping configuration. Approximation achieves higher performance without changing the running power trends for \textit{Welch} based EEG applications. Hence, higher energy efficiency can be achieved at a higher level of approximation.
Figure \ref{fig.3d_600.Apx} presents a 3D model of power-performance vs. accuracy trade-offs for all three applications. We show 12x8 data points representing accurate and approximate configurations at both CPU clusters. The increasing size of the data marker shows the increasing number of mapping cores. The data points are closer to each other for the LITTLE cluster, and the overlap decreases when the applications run on multiple big cores. The overlapping space is important for design prospects showing that each application can have similar power, performance, and accuracy trends for different configurations.
\begin{figure*}[t]
\vspace{-3pt}
\centering
\includegraphics[width=0.99\textwidth]{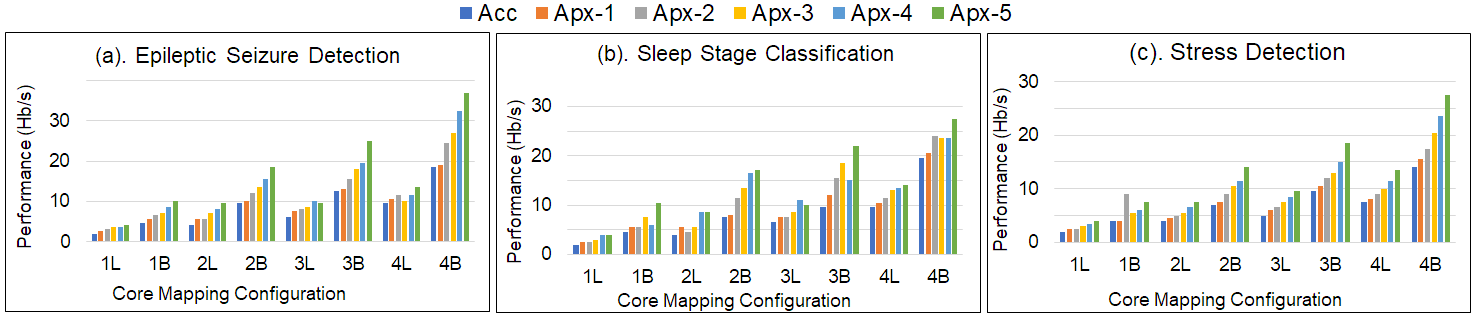}
\caption{Performance characteristics at 1000 Hz for (a) Epilepsy, (b) Sleep, (c) Stress.}
\label{fig.Perf_1000.Apx}
\vspace{-6pt}
\end{figure*}

\begin{figure*}[t]
\vspace{-3pt}
\centering
\includegraphics[width=0.99\textwidth]{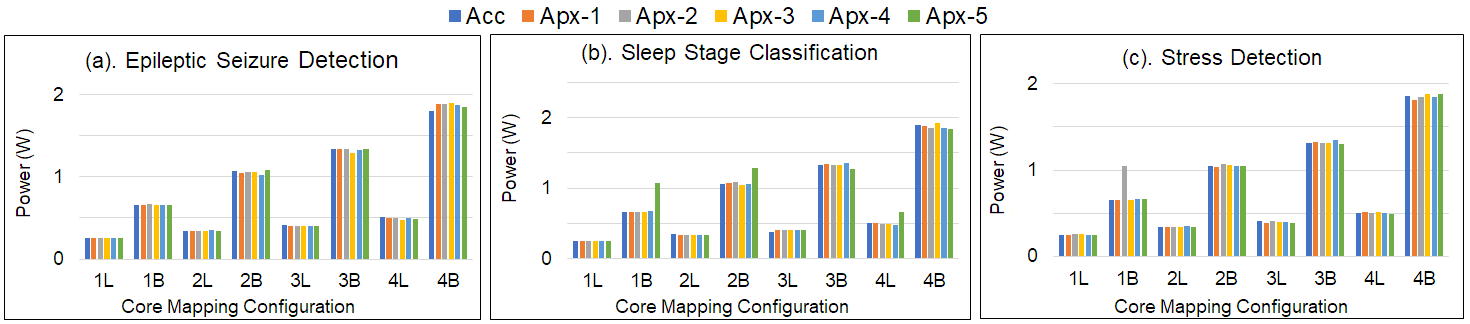}
\caption{Power characteristics at 600 Hz for (a) Epilepsy, (b) Sleep, (c) Stress.}
\label{fig.Power_1000.Apx}
\vspace{-6pt}
\end{figure*}

\begin{figure*}[t]
\vspace{-3pt}
\centering
\includegraphics[width=0.99\textwidth]{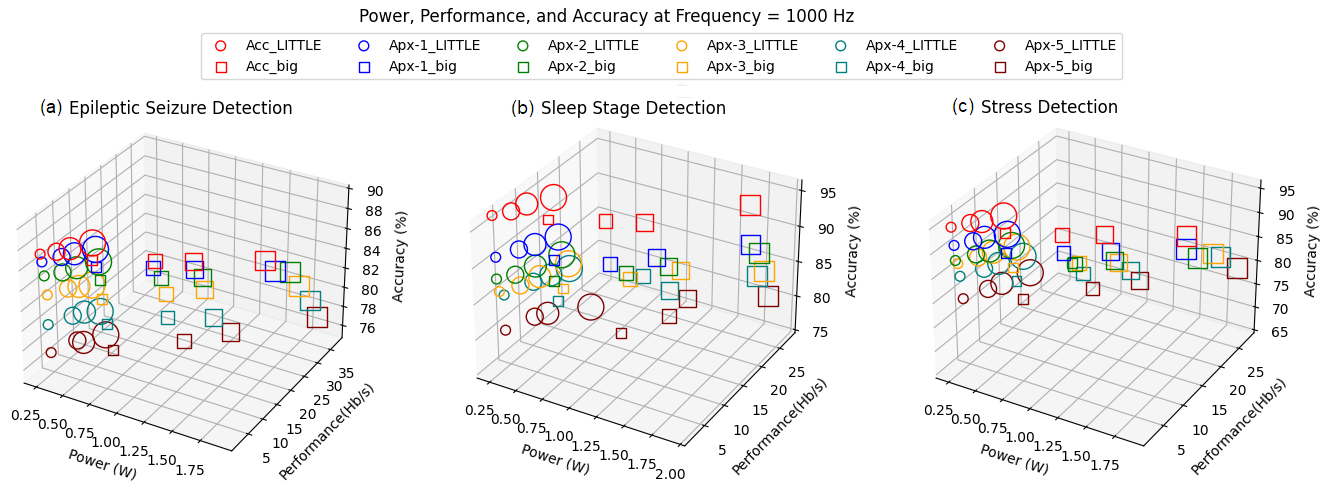}
\caption{Accuracy trade-off vs power-performance characteristics at 1000 Hz for (a) Epilepsy, (b) Sleep, (c) Stress.}
\label{fig.3d_1000.Apx}
\vspace{-12pt}
\end{figure*}
For Epileptics Seizure Detection, the performance characteristics are nearly the same between four LITTLE cores and one big core. Hence, there is an overlapping space between the data point of the LITTLE cluster and one big core.
For Sleep Stage Detection, the power consumption is nearly the same when the application is mapped on the LITTLE cluster. There is a clear difference in the accuracy gap between the accurate configuration and level-5 approximation. The data point for the remaining approximation levels is closer.
For Stress Detection, there is a clear distance between the data point of the big cluster. However, the remaining core mapping configurations of the LITTLE cluster or one big core have an overlapping space.

\textbf{Frequency Level 2.} In this configuration, we ran all three applications at 1000Hz. Figure \ref{fig.Perf_1000.Apx} shows the performance trends at eight different core mapping configurations. The performance gain at increasing approximation levels can be seen in each mapping configuration. Figure \ref{fig.Power_1000.Apx} represents the power trends for all the applications at the 1000Hz frequency level. For the same number of cores, the running power on the big cluster is significantly higher than on the LITTLE cluster. Figure \ref{fig.3d_1000.Apx} shows the 3-D comparison between power, performance, and accuracy of all three applications at different approximation levels.
\begin{figure*}[t]
\vspace{-3pt}
\centering
\includegraphics[width=0.99\textwidth]{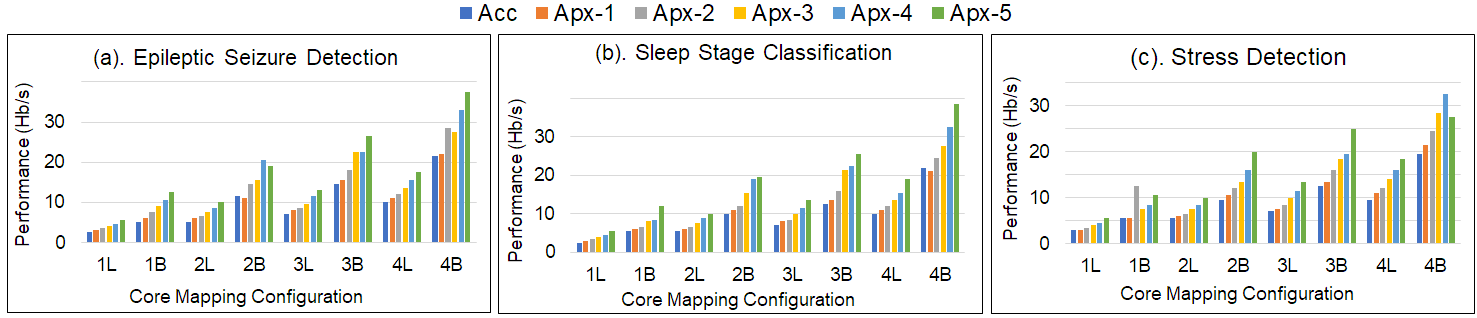}
\caption{Performance characteristics at 1400 Hz for (a) Epilepsy, (b) Sleep, (c) Stress.}
\label{fig.Perf_1400.Apx}
\vspace{-6pt}
\end{figure*}

\begin{figure*}[t]
\vspace{-3pt}
\centering
\includegraphics[width=0.99\textwidth]{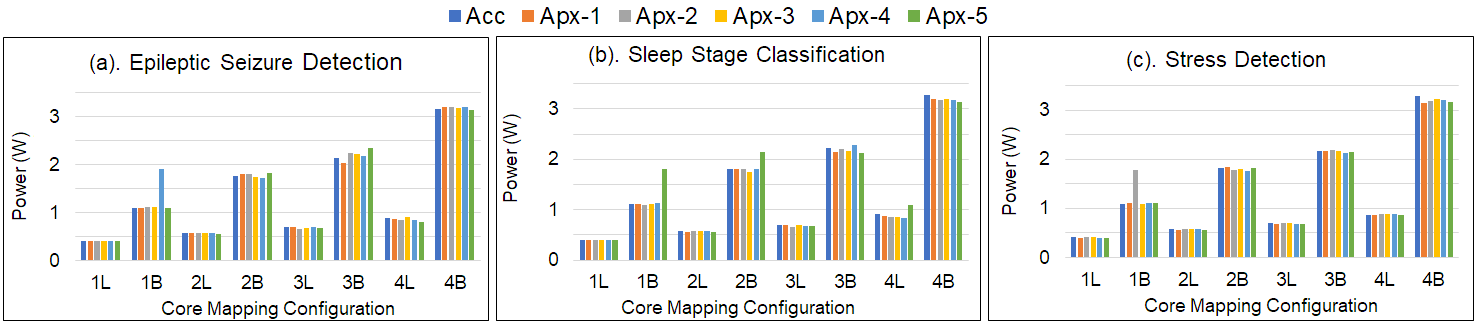}
\caption{Power characteristics at 1400 Hz for (a) Epilepsy, (b) Sleep, (c) Stress.}
\vspace{-6pt}
\label{fig.Power_1400.Apx}
\end{figure*}

\begin{figure*}[t]
\vspace{-3pt}
\centering
\includegraphics[width=0.99\textwidth]{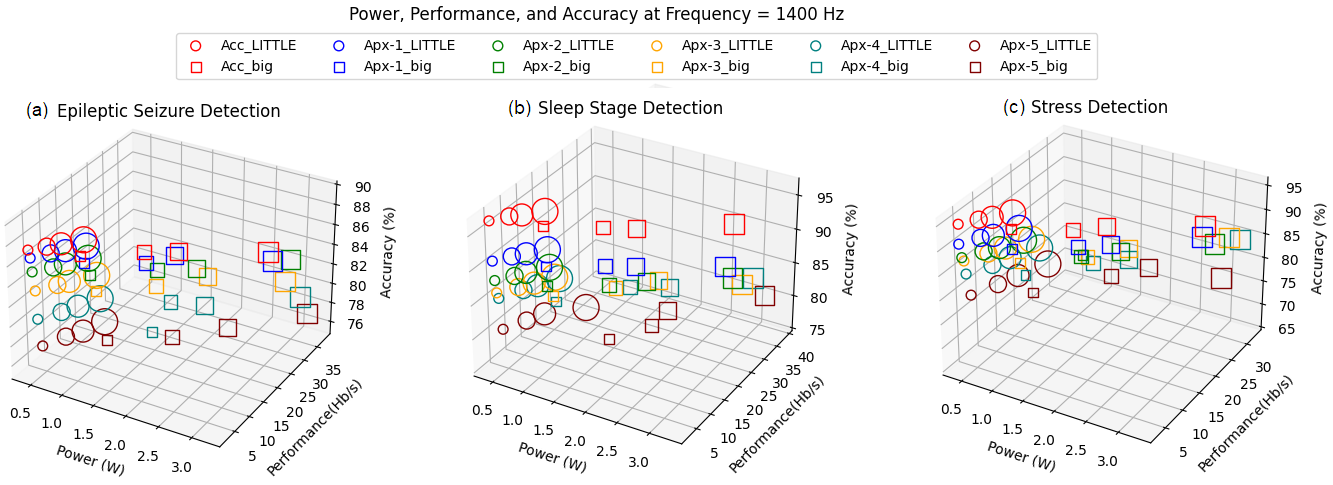}
\caption{Accuracy trade-off vs power-performance characteristics at 1400 Hz for (a) Epilepsy, (b) Sleep, (c) Stress.}
\vspace{-12pt}
\label{fig.3d_1400.Apx}
\end{figure*}

\noindent In Epileptic Seizure Detection, there are larger gaps between data points in the LITTLE cluster. The application is getting higher performance at four LITTLE cores than one big core. In contrast, the power at one big core is significantly higher than any configuration on LITTLE.
In Sleep Stage Detection, the accuracy gaps are significantly large between the accurate and approximation level-5. Moreover, there is a non-linear increase in performance between different approximation levels at each core mapping configuration.
In Stress Detection, the accuracy gaps are low between different approximation levels. Moreover, the power between different configurations on the LITTLE cluster is almost identical. However, the performance is nearly the same on the core mapping configuration of one big core and two LITTLE cores. Similarly, we can see the same performance trends between two big and three LITTLE cores. However, the power consumption at the LITTLE cluster is significantly lesser than in the big cluster.
\noindent\textbf{Frequency Level 3.} We ran all three EEG applications on eight different core mapping configurations at 1400 Hz. Figure \ref{fig.Perf_1400.Apx} shows the comparison of performance at different approximation levels for the applications. Similar two the previous scenarios, the performance trends show higher performance on big clusters than on LITTLE for an equal number of cores. Figure \ref{fig.Power_1400.Apx} shows the power trends for all three applications. The highest power at four big cores mapping is crossing 3W. However, the general trends show that the running power remains almost similar for each level of approximation at a given core mapping configuration. Figure \ref{fig.3d_1400.Apx} presents the complete comparison of power, performance, and frequency. Here, the stress detection application has a closer overlap between data points. All the applications show a greater distance between data points when running on multiple big cores.
In Epileptic Seizure Detection, the power consumption increases with an increasing number of cores in both clusters. Hence, the overlapping gap is lesser than in the previous configurations. The performance at four LITTLE cores is higher than one big core at a lesser power consumption cost.
In Sleep Stage Detection, the power consumption in the big cluster is significantly higher than the LITTLE cluster for any given number of cores. The performance at four LITTLE cores is nearly identical to two big cores at a lower power consumption cost.
In Stress Detection, the power consumption trend is still similar, and a visible gap appears between the data point of the LITTLE cluster and the big cluster. However, the performance at four LITTLE cores is higher than one big core and nearly identical to two big cores. However, the power consumption cost at any core mapping configuration of the LITTLE cluster is significantly lesser than the big cluster.

\noindent\textbf{Discussion:} We evaluated the power, performance, and accuracy trends for all three EEG applications on the Odroid XU-3 board. We could observe data configurations for each application where the characteristics were overlapping or very close to each other. We saw the general trend that the LITTLE cluster is more energy efficient. Still, it showed better performance at a lower power cost than single-core configurations of the big cluster. Moreover, we saw the general trend in the approximation that the performance of each application was linearly increasing with the increase of approximation level at any core mapping configuration. The power consumption was the same at any approximation level for a given core mapping configuration. Hence, for constant running power, increasing the approximation level provides better performance, providing better energy efficiency.

\section{Conclusion}
We characterized three EEG applications, including Epileptic Seizure Detection, Sleep Stage Classification, and Stress Detection on the Odroid XU3 platform. We investigated each application's power, performance, and accuracy trends at different cores of each CPU cluster. We compared the power-performance metrics against accuracy trade-offs at five different approximation levels. We exploited the error resilience nature of the Welch method to implement application-level approximation.

\section*{Acknowledgements}
We gratefully acknowledge the funding from European Union's Horizon 2020 Research and Innovation programme under the Marie Sk\l{}odowska Curie grant agreement No. $956090$ (APROPOS: Approximate Computing for Power and Energy Optimisation)

\bibliographystyle{plain}
\bibliography{reference}

\end{document}